\documentclass{aa}  

\usepackage{graphicx}
\usepackage{txfonts}

\def\be{\begin{equation}}
\def\ee{\end{equation}}
\def\bi{\begin{itemize}}
\def\ei{\end{itemize}}
\def\tt{\texttt}

\usepackage{lscape}  

\begin{document} 

\title{ISIS: a new N-body cosmological code with scalar fields based on RAMSES}
\subtitle{Code presentation and application to the shapes of clusters}
\author{Claudio Llinares \inst{1, 2}
  \and
  David F. Mota \inst{1}
  \and
  Hans A. Winther \inst{1}
}

\institute{$^1$Institute for Theoretical Astrophysics, University of Oslo, P.O. Box 1029 Blindern, N-0315 Oslo, Norway, \\
$^2$Hamburger Sternwarte, Gojenbergsweg 112, 21029 Hamburg, Germany}
\date{Submitted June 2013}

\titlerunning{ISIS: a new N-body cosmological code with scalar fields}
\authorrunning{C. Llinares et al.}

\abstract{Several extensions of the standard cosmological model include scalar fields as new degrees of freedom in the underlying gravitational theory.  A particular class of these scalar field theories include screening mechanisms intended to hide the scalar field below observational limits in the solar system, but not on galactic scales, where data still gives the freedom to find possible signatures of their presence. To make predictions to compare with observations coming from galactic and clusters scales (i.e. in the non-linear regime of cosmological evolution), cosmological N-body simulations are needed, for which codes that can solve for the scalar field must be developed. We present a new implementation of scalar-tensor theories of gravity that include screening mechanisms. The code is based on the already existing code RAMSES, to which we have added a non-linear multigrid solver that can treat a large class of scalar tensor theories of modified gravity. We present details of the implementation and the tests that we made to the code.  As application of the new code, we studied the influence that two particular modified gravity theories, the symmetron and $f(R)$ gravity, have on the shape of cluster sized dark matter haloes and found consistent results with previous estimations made with a static analysis.}


   \maketitle

\section{Introduction}

While the standard model for cosmology, $\Lambda$CDM, is widely accepted as a possibly valid explanation for reality, there are several issues that are not fully understood on galactic scales, which give theoreticians the chance to consider extensions to the model. Furthermore, the tension found by the Planck collaboration \citep{sigma8_planck} between $\sigma_8$ and $\Omega_m$, likewise puts the model in check on cosmological scales. The same collaboration also confirmed the already known anisotropy of the CMB \citep[][]{{2004ApJ...605...14E}, {2009ApJ...704.1448H},{planck_param}}, which is difficult to explain within the standard model.  Between the possible extensions to the model, there is the idea of modifying the gravitational theory. Several alternative theories exist \citep[][]{{2012PhR...513....1C}, {2012arXiv1206.1225A}}, all of which include extra degrees of freedom in the form of scalar, vectors, and even tensor fields.  To test these models in the non-linear regime of structure formation by using large surveys, such as the upcoming Euclid \citep[][]{2011arXiv1110.3193L} and LSST \citep[][]{2009arXiv0912.0201L} surveys, precise and accurate predictions are needed, for which numerical simulations must be performed.  Within the standard context, there are several algorithms and very well tested codes that are known to give consistent results. For models beyond $\Lambda$CDM, however, the situation is still not settled, and only a few codes exist per alternative gravitational model \citep[e.g.][]{{2008arXiv0809.2899L}, {2008PhRvD..78l3523O}, schmidt_dgp_code, baldi_coupled_quintessence_code, llinares_thesis, zhao_baojiu_forf_code, {2012JCAP...01..051L}, baldi_de_sim, {2012MNRAS.422.1028B}, 2012JCAP...10..002B, 2013MNRAS.436..348P, nbody_chameleon, dgp_code_durham,li1,li2}.

 N-body techniques are crucial in the build up of predictions, and thus, they should be developed independently by more than one research group.  In an effort to extend the existing library of codes and give strength to the results by showing that they are stable when changing underlying approximations and implementation details, we present here a new implementation of scalar-tensor modified gravity theories in the code \tt{RAMSES} \citep[][]{2002A&A...385..337T}.

The set of models we have focussed on are scalar-tensor models that were originally designed as explanation for dark energy and that include screening mechanisms, which are induced by non-linearities in the equation of motion for the scalar field. The dominant dynamical effects appear in this models through the inclusion of a fifth force in addition to the gravitational force. We are interested in the effects of this fifth-force in the evolution of large scale structure and the formation of dark matter haloes in particular. The code that we present here must be taken not as definitive, but as a starting point for more complex simulations including hydrodynamics and different types of couplings, including even non-universal couplings to the different matter species found in our Universe.  

For a large class of scalar-tensor theories, one finds the following equation of motion for the metric perturbation $\Phi$, the scalar field $\phi$, and the positions $\mathbf{x}$ of the N-body particles: 
\begin{align}
\label{eqs_general_1}
& \nabla^2 \Phi = \frac{3}{2}\frac{\Omega_m H_0^2}{a} \delta,\\
\label{eqs_general_2}
& \nabla^2\phi = S(\phi, \rho, a),\\
\label{eqs_general_3}
& \ddot{\mathbf{x}} + 2 H \dot{\mathbf{x}} + \frac{1}{a^2} \nabla\Phi + \mathbf{g}(\phi, \nabla\phi, a) = 0,
\end{align}
where $S$ and $\mathbf{g}$ are model specific functions.  These equations are the outcome of canonical scalar tensor theories, and they can also be applied to the study of $f(R)$ theories, which can be recast as a scalar-tensor theory.  The hard part of trying to solve equations (\ref{eqs_general_1}) to (\ref{eqs_general_3}) consists of changing the original linear multigrid solver of \tt{RAMSES} to a non-linear one. In our implementation, the functions $S$ and $\mathbf{g}$ are not hard-coded, but left as free functions, which increases the flexibility regarding the models that can be simulated.

Equations (\ref{eqs_general_1}) to (\ref{eqs_general_3}) are the consequence of assuming the quasi-static approximation (i.e. time derivatives in the fields were neglected).  \citet[][]{2013PhRvL.110p1101L} presented a non-static solver, which raises the question about the validity of this approximation.  However, one has to take into account that non-static simulations are very costly since they track very rapid oscillations of the scalar field in time, which implies the use of very short time steps. Furthermore, the frequency of the oscillations grows when going to small scales, which will imply even higher requirements when increasing resolution.  While \citet[][]{2013PhRvL.110p1101L} proved that some of the properties of static and non-static solutions can disagree, the static simulations are still needed to calibrate non-static methods and to work out observables that are not affected by the oscillations.  In other words, while the static approximation could give biased predictions for a number of (still undetermined) observables, it gives much higher flexibility regarding resolution and will continue to be used in the future.

The paper is organised as follows.  Section \ref{section:general_equations} presents the general set of equations we intend to solve, while section \ref{section:method} describes the discretisation in detail that we use as well as our implementation of the non-linear multigrid algorithm in the code \tt{RAMSES}. In section \ref{section:specific_models} we give the model-specific equations for two different modified gravity theories.  Tests that we have made of the code are presented in section \ref{section:tests}, and section \ref{section:application} shows an application of the code where we calculate the impact that scalar fields have on the shape of dark matter haloes. Finally, conclusions and discussion are given in section \ref{section:conclusions}. 

\section{The equations for generic scalar fields}
\label{section:general_equations}
We are interested in running simulations with models defined by the following action:
\be
S = \int \sqrt{-g} \left[ R - \frac{1}{2}\nabla^a\phi \nabla_a \phi - V(\phi)\right] d^4x + S_M(\tilde{g}_{\mu\nu}, \psi)
\label{action_scalar_tensor}
\ee
where the Einstein and Jordan frame metrics ($g_{\mu\nu}$ and $\tilde{g}_{\mu\nu}$) are related according to
\be
\tilde{g}_{\mu\nu} = A^2(\phi) g_{\mu\nu}.
\ee
The equation of motion for the scalar field that results from this Lagrangian is
\be
\Box\phi = V_{,\phi} - A_{,\phi} T,
\ee
where $T$ is the trace of the Einstein frame energy momentum tensor.  To be able to introduce this equation into the code in the cosmological context, we need to specify the metric
\be
ds^2 = -(1+2\Phi)dt^2 + a^2 (1-2\Phi)(dx^2+dy^2+dz^2),
\ee
which is a flat Friedmann-Lema$\hat{\text{i}}$tre-Robertson-Walker metric with scalar perturbations. With this metric, the equation of motion in the quasi-static limit reads as
\be
\frac{1}{a^2}\nabla^2\phi = V_{,\phi} + A_{,\phi} \rho \equiv S(\rho, \phi),  
\label{canonical_equation}
\ee
where $\rho$ is the matter density and $S$ the source-term shown in Eq. (\ref{eqs_general_2}).

In certain models, it is convenient for numerical reasons to redefine the scalar field
\be
\phi = j(u),
\label{change_non_canonical}
\ee
where the function $j$ is chosen such that it fixes the sign of the scalar field to be unique and at the same time, it reduces extreme gradients that the scalar field could have.  Typical choices for $j$ are power laws or exponential function. The field equation of the new field $u$ is
\be
\nabla\cdot \left[b(u)\nabla u\right] = S(\rho, u),
\label{non_canonical_equation}
\ee
where 
\be
b = \frac{dj}{du}.
\ee
Thus, even though we start with a canonical scalar field, the equation we end up trying to solve will not be canonical in many cases. Our code must therefore be able to solve non-canonical equations, and we discuss how this is done in the next section.  Naturally, since the redefinition is usually non-linear, it must be made after switching to a dimensionless field, which we describe in section \ref{section:specific_models} when specifying the models.

The evolution of the matter component is found by discretizing the density field with particles and finding their free trajectories, which are given by the geodesics equation.  By taking the terms that involve the scalar field in the action into account, one obtains the following modified geodesics:
\be
\ddot{\bf x}+ 2H\dot{\bf x} +\frac{1}{a^2}\nabla\Phi + \frac{1}{a^2}\nabla\log A(\phi) = 0, 
\label{geodesics_general}
\ee
where we also neglected non-static terms.

\section{Implementation of scalar fields in Ramses}
\label{section:method}

Our code is a modification of the open source N-body code \tt{RAMSES} \citep[][]{2002A&A...385..337T}, to which we added a non-linear implicit solver to consider the equation of motion of the scalar field in its static approximation.  Information about the original multigrid linear solver that we employed as starting point can be found in \citet[][]{2011JCoPh.230.4756G}.  The set of variables for which the code is written in are defined in \citet[][]{1998MNRAS.297..467M}.  

In brief, the \tt{RAMSES} code is an N-body particle mesh code that also includes a Godunov solver to treat the evolution of baryons.  The gravitational forces are calculated as spatial derivatives of the gravitational potential that is previously calculated on a grid. We give here a few words regarding the original linear Poisson solver and present the modifications that are necessary for including the non-linear equation for the scalar field in its canonical and non-canonical forms given by Eqs. (\ref{canonical_equation}) and (\ref{non_canonical_equation}) respectively.

\subsection{Poisson's equation}

In the standard gravity case, the code \tt{RAMSES} solves the following equation for the gravitational potential: 
\be
\nabla^2 \Phi = \frac{3}{2}\frac{\Omega_m H_0^2}{a} \delta = S(\delta), 
\ee
where $\delta=\delta\rho/\rho_0$ is the over-density.  The equation is solved by discretizing the differential operator $\nabla^2$ on a cartesian grid and applying a Gauss-Seidel iteration scheme to the resulting algebraic equation. Multigrid techniques are implemented to accelerate the convergence of the method.  

The Laplacian is discretized by using a standard second-order formula
\be
\nabla^2\Phi= \frac{(\Phi_{i+1}+\Phi_{i-1}+\Phi_{j+1}+\Phi_{j-1}+\Phi_{k+1}+\Phi_{k-1}) - 6\Phi}{h^2}, 
\label{disc_lap}
\ee
where we show only sub-indexes with values different than $(i,j,k)$, which is the notation that we use throughout the paper.  The standard code solves the previous equation by means of an explicit iterative method that starts from an initial guess for the potential.  During each iteration step, the potential is changed in an explicit way according to
\begin{multline}
\Phi= \left\{(\Phi_{i+1}+\Phi_{i-1}+\Phi_{j+1}+\Phi_{j-1}+\Phi_{k+1}+\Phi_{k-1}) - \right.\\
\left. h^2 S(\delta)\right\} / 6.
\label{explicit}
\end{multline}
To speed up the convergence, the solver includes multigrid relaxation.  In brief, a two-level algorithm is as follows. Given an approximation for the solution $\phi^k$ the code solves the equation
\be
\nabla^2\delta\phi^k = R(\epsilon^k)
\label{coarse}
\ee
for a coarse grid correction $\delta\phi$,  where $R$ is a restriction operator, and $\epsilon$ the fine grid residual, which is defined as
\be
\epsilon^k=\nabla^2\phi^k - \rho.
\ee
Once a fixed number of Gauss-Seidel iterations is made for solving Eq.~(\ref{coarse}), a correction is applied to the original (fine grid) solution in the following way:  
\be
\bar{\phi}^k = \phi^k + P(\delta\phi^k),
\ee
where $P$ is a prolongation operator that moves the information from the coarse to the fine grid.  The standard \tt{RAMSES} code not only uses two levels for the iterations, but also several levels within a $V$ scheme (i.e. iterations are made starting from the finest grid down to the coarsest one and coming back up making corrections in every level).  More details on multigrid relaxation and its generalisation to non-linear equations can be found, for instance, in \citet[][]{Brandt77}, \citet[][]{Wesseling92}, or \citet[][]{Trottenberg}.

\subsection{Extending the original solver to non-linear equations (full approximation storage)}

Our solver for the scalar field is based on the original Poisson's solver, however, since the source of the scalar field equation will for many models have a non-linear dependence on the scalar field, the previous procedure can no longer be used. Our generalisation of the multigrid scheme is based on the full approximation storage (FAS) algorithm \citep[e.g.][]{Brandt77}.  In this case, the equation solved in the coarse grids is no longer a correction to the solution, but the solution itself.  The original equations for the scalar field, Eq.~(\ref{canonical_equation}) or (\ref{non_canonical_equation}), can be written in the form
\be
L(\phi, \rho) = \Sigma, 
\ee
where the operator $L$ is given in the canonical case by
\be
L(\phi,\rho) = \nabla^2\phi - S(\rho, \phi)\\
\ee
and by
\be
L(u,\rho) = \nabla\cdot \left[b(u)\nabla u\right] - S(\rho,u)
\ee
in the non-canonical one.  The new source $\Sigma$ is zero in the fine grid and has the following expression in the coarse grids,
\be
\Sigma = -R[\epsilon(\phi,\rho)] + \epsilon(R\phi, R\rho),
\ee
where $R$ is a restriction operator that moves information from the original fine grid to the coarse one, and the residual $\epsilon$ is defined as
\be
\epsilon(\phi,\rho) = L(\phi,\rho) - \Sigma.
\ee
The source $\Sigma$ is only calculated when jumping from a fine to a coarse grid.

The Gauss-Seidel iterations that are needed to improve the solution of the discretized equation are performed in an implicit way
\be
\bar{\phi} = \phi - \frac{L(\phi,\rho)-\Sigma}{\partial L(\phi,\rho)/\partial\phi}.
\ee
This expression can be derived as one step of a Newton-Ralphson scheme applied to the solution of the equation
\be
L(\phi,\rho)-\Sigma = 0,
\ee
and assuming that 
\be
\frac{\partial \Sigma}{\partial\phi} = 0,
\ee
which was found to be a good approximation.

The non-linear multigrid algorithm can be implemented in the same way in both the canonical and non-canonical cases. The only difference between the two equations is the discretization formula used, which has to be written explicitly for each differential operator in the uniform part of the grid and in the boundaries of the refinements.  We present these details in the following sections.

\subsection{Discretization of the canonical equation}

The canonical Eq.~(\ref{canonical_equation}) is discretized using standard second-order formulas as in the original \tt{RAMSES} code, but applied to the scalar field $\phi$ instead of the gravitational potential $\Phi$. The derivative of the discretised differential operator that are needed for the implicit iterations is given by
\be
\frac{dL}{d\phi} = -\frac{6}{h^2} - \frac{dS}{d\phi}, 
\ee
where we have taken into account that the source $S$ in Eq.~(\ref{canonical_equation}) is now a function of both the matter density and the scalar field $\phi$.

\subsubsection{Discretization near the boundaries of the refinements (canonical case)}

The standard \tt{RAMSES} code includes adaptive mesh refinements, which means that the resolution of the grid is not uniform in space, but is increased in regions of interest.  The decision to refine a cell in the fine grid is given by some specific criteria, which could be, for instance, a density threshold. The nodes that lie on the border of the refinement patches lack one or more neighbours and thus, the discretization formula must be modified to take this into account.  We briefly describe the workaround to this problem that is implemented in the standard code and in our extension to the non-linear case.  We give a simplified discussion of the canonical case, but describe the complete algorithm in the following section when showing the discretization for the non-canonical equation.

The reconstruction method included in the standard code is based on the fact that the standard formula for the Laplacian, Eq.~(\ref{disc_lap}), can be rewritten as
\be
\frac{\partial^2 \phi}{\partial x^2} = \left. \left(\frac{\phi_{i+1}-\phi}{h}-\frac{\phi-\phi_{i-1}}{h}\right) \middle/ h \right. ,
\ee
where we show derivatives in only one dimension.  To explain the procedure, we assume, for instance, that we want to calculate the Laplacian in a cell $(i,j,k)$, which only lacks the neighbour $(i+1,j,k)$ (i.e. the node $(i,j,k)$ is the last one of a refinement to the right in $x$ direction).  The derivative in the $x$ direction is calculated by using information that must be reconstructed in the boundary of the refinement, which we call $\phi_{i+1/2}$.  The modified expression for this derivative is
\be
\frac{\partial^2 \phi}{\partial x^2} = \left. \left(\frac{\phi_{i+1/2}-\phi}{h/2}-\frac{\phi-\phi_{i-1}}{h}\right)\middle/h \right., 
\ee
where $\phi_{i+1/2}$ is obtained in the first place by moving the information from the next neighbour coarse node to the neighbour non-existing node $\phi_{i+1}$.  Next, the information is moved from there to the border of the refinement (giving us $\phi_{i+1/2}$).  When the addition for the 3D Laplacian is made, we obtain the following discretization formula:
\be
\nabla^2\phi= \frac{(2\phi_{i+1/2}+\phi_{i-1}+\phi_{j+1}+\phi_{j-1}+\phi_{k+1}+\phi_{k-1}) - 7\phi}{h^2}.
\label{gen_lap}
\ee
In the general case, the number 7 will be substituted by the number of boundaries $n$ that the node $(i,j,k)$ contains.  See \citet[][]{Gibou_asecond-order-accurate} for more details on alternative discretization formulas.

The equation for the explicit iterations, Eq.~(\ref{explicit}), is then modified and written as
\begin{align}
\nonumber
\phi_i = \frac{1}{7}\left\{(\phi_{i-1}+\phi_{j+1}+\phi_{j-1}+\phi_{k+1}+\phi_{k-1}) - \right.\\
\left. - h^2\left[\rho-\frac{2\phi_{i+1/2}}{h^2}\right]\right\}. 
\label{rhs}
\end{align}
The term in square brackets in this expression can be seen as a modified source and calculated before any iteration is made, which is the way the algorithm is implemented in the original code. However, in the non-linear case, the previous expression is not valid because the Laplacian and the solution have a non-linear dependence on $\phi$. In the modified code, the complete Laplacian is calculated at each iteration step. This can be done by rearranging terms in the discretized Laplacian in the following way:
\be
\nabla^2\phi= \frac{2\phi_{i+1/2}}{h^2} + \frac{\phi_{i-1}+\phi_{j+1}+\phi_{j-1}+\phi_{k+1}+\phi_{k-1} - 7\phi}{h^2}.
\label{symm_lap}
\ee
The first term can be now calculated only once, before any iteration is made.  The second term must be calculated at every Gauss-Seidel iteration step.

\subsection{Discretization of the non-canonical equation}
\label{discretiz_non_canonical}

The discretized differential operator $L$ that corresponds to the non-canonical Eq.~(\ref{non_canonical_equation}) can be written as
\begin{align} 
\nonumber  \nabla & \cdot \left[b(u)\nabla u \right] = \\
\nonumber & b_{i+1/2}u_{i+1} + b_{i-1/2}u_{i-1} + b_{j+1/2}u_{j+1} + \\
\nonumber & b_{j-1/2}u_{j-1} + b_{k+1/2}u_{k+1} + b_{k-1/2}u_{k-1} -\\
& (b_{i+1/2}+b_{i-1/2}+b_{j+1/2} + b_{j-1/2}+b_{k+1/2}+b_{k-1/2})u, 
\label{disc_non_canonical}
\end{align}
where we used the same notation as in previous paragraphs (i.e. we show only sub-indexes with values different than $(i,j,k)$).  By assuming $b(u)=1$, we recover the standard formula for the Laplacian.  The values of $b$ at the faces of the nodes are calculated by linear interpolation 
\begin{align}
& b_{i\pm1/2} = \frac{b(u)+b(u_{i\pm1})}{2} \\
& b_{j\pm1/2} = \frac{b(u)+b(u_{j\pm1})}{2} \\
& b_{k\pm1/2} = \frac{b(u)+b(u_{k\pm1})}{2}.
\end{align}
The derivative of Eq.~(\ref{disc_non_canonical}) needed for the implicit scheme is given by
\begin{align}\label{dbeq}
& \frac{\partial}{\partial u} \left\{ \nabla \cdot \left[b(u)\nabla u \right]\right\}= \\
& -\frac{(b_{i+1/2}+b_{i-1/2})}{h^2} + \frac{db}{du}\frac{(u_{i+1}+u_{i-1}-2u)}{h^2} - \nonumber\\
&-\frac{(b_{j+1/2}+b_{j-1/2})}{h^2} + \frac{db}{du}\frac{(u_{j+1}+u_{j-1}-2u)}{h^2} - \nonumber\\
&-\frac{(b_{k+1/2}+b_{k-1/2})}{h^2} + \frac{db}{du}\frac{(u_{k+1}+u_{k-1}-2u)}{h^2} - \nonumber\\
&- \frac{\partial S(u,\rho)}{\partial u}.
\end{align}

\subsubsection{Discretization near the boundaries of the refinements (non-canonical case)}

\begin{figure}
\includegraphics[width=1\columnwidth]{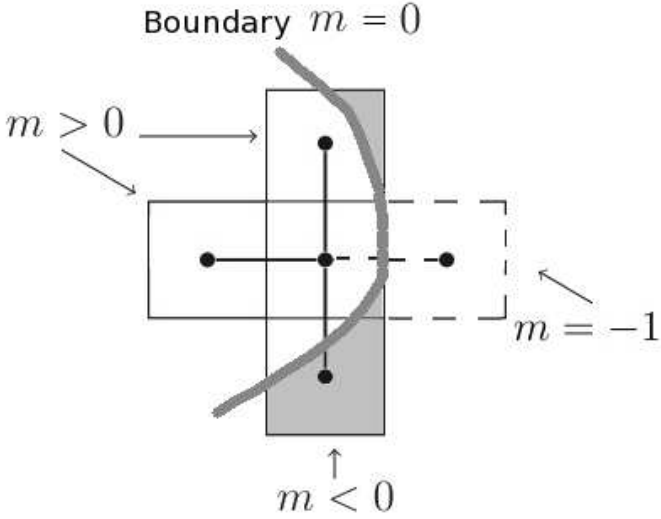}
\caption{A five points stencil (2D) located close to the boundary. The central cell has two neighbours that are not masked $m>0$ and two cells that are masked $m<0$. One of these cells is completely masked $m=-1$ meaning that it does not exist in the memory, and its field value must be interpolated from the coarse grid. For the cells that are only partly masked $-1<m<0$, the field value is determined from equations (\ref{bpsi}) and (\ref{intpsi}). The cells with $m>0$ are treated as normal cells even though the boundary might cross some part of the cell. }
\label{fig:stensil}
\end{figure}

When the multigrid algorithm is applied not only to the domain grid (which is uniform), but also to the refined regions, we face the problem that the boundary of a node that belongs to a refinement does not always correspond to the boundary of its corresponding coarse nodes.  The issue was solved in the standard \tt{RAMSES} code by introducing a mask function in the grids to indicate which of their nodes belongs to a given refinement or does not.  Inner cells (i.e. cells that are not in the boundary) are initialised with a mask function with value $m_{i,j,k} = 1$ and outer cells with a mask $m_{i,j,k} =-1$. The boundary is defined at the position where the interpolated value of the cell-centred mask crosses zero. In the finest grid, the boundaries are positioned at the faces of the outer cells. On the corresponding coarser grids, the mask is calculated by applying the restricting operator to the mask. The mask values must be understood as follows. Cells with positive mask value have their centre inside the refinement. In that case, one can use the inner discretisation formulas described above. Cells with negative mask value exists as refined cells, but have their centre outside the boundary. Finally, cells where the mask takes the value -1 are completely outside the refinement (i.e. they do not exist).  Figure \ref{fig:stensil} shows an example of the distribution of masks in a node that belongs to a 2D stencil and is close to the boundary of the refinement.

To deal with the cases where we have masked cells, the differential operator is redefined for cells ($i,j,k$) close to the boundary whenever one of the six neighbouring cells has negative mask value. The information in cells with $m<0$ is replaced by a ghost value linearly interpolated between cell $i$ and the boundary value. This ghost value depends explicitly on cell $(i,j,k)$ and the boundary condition, and with this we make sure that the boundary remains at the same location to second-order accuracy when one goes from fine to coarse levels in the multigrid hierarchy.

To explain the procedure, we assume for simplicity that only the $(i+1,j,k)$'th cell is masked ($m_{i+1,j,k}<0$). By linear interpolation, we find that the distance from the cell $(i,j,k)$ to the place where the mask-value crosses $m=0$ (i.e. the position of the boundary) is at
\begin{align}
\omega = \frac{m_{i+1,j,k}}{m_{i+1,j,k}-m_{i,j,k}}
\end{align}
times the distance $x_{i+1,j,k}-x_{i,j,k}$. The boundary value $u_{\#}$ of the scalar field can then be found as
\begin{align}\label{bpsi}
u_{\#} = (1-\omega)u^{\rm Int}_{i+1,j,k} + \omega u_{i,j,k}
\end{align}
where $u^{\rm Int}_{i+1,j,k} = u_{i+1,j,k}^{\rm Pre}$ if $-1<m_{i+1,j,k}\leq 0$ and if $m=-1$ (if the cell does not exist) we find  $u^{\rm Int}_{i+1,j,k}$ by interpolating it from the coarse grid. The value $u_{i+1,j,k}^{\rm Pre}$ in previous expresion is the value of the cell $(i+1,j,k)$ before we make the Gauss-Seidel iterations (at the time when the boundary value $u_{\#}$ is defined).

The ghost value for $u_{i+1,j,k}$ can then be found from
\begin{align}\label{intpsi}
u^G_{i+1,j,k} = u_{\#}\left(1- \frac{m_{i+1,j,k}}{m_{i,j,k}}\right)  + \frac{m_{i+1,j,k}}{m_{i,j,k}} u_{i,j,k}.
\end{align}
We can combine Eqs.~(\ref{bpsi}) and (\ref{intpsi}) to get the following equation for the value of $u$ in the masked cell
\begin{align}\label{ueq}
u^G_{i+1,j,k} = u^{\rm Int}_{i+1,j,k} +  \frac{m_{i+1,j,k}}{m_{i,j,k}}(u_{i,j,k}-u_{i,j,k}^{\rm Pre}).
\end{align}

Owing to the non-linearity of the differential operator, it is not possible to include these modifications in the source term before making the Gauss-Seidel iterations. We have chosen to solve this by storing the value of $u_{i,j,k}^{\rm Pre}$ for the boundary cells so that we can reconstruct the ghost value for $u_{i+1,j,k}$ whenever needed during the Gauss-Seidel iterations.

To complete the discretization in the nodes that are close to the boundary, we need to define a consistent value for $b_{i+1,j,k}$ and $c_{i+1,j,k} \equiv \left(\frac{db}{du}\right)_{i+i,j,k}$. One way to do this, proposed in \citet[][]{Gibou_asecond-order-accurate}, is to use
\begin{align}
b_{\#} &= (1-\omega)b^{\rm Int}_{i+1,j,k} + \omega b_{i,j,k},\\
b_{i+1,j,k}^G &= b^{\rm Int}_{i+1,j,k} +  \frac{m_{i+1,j,k}}{m_{i,j,k}}(b_{i,j,k} - b_{i,j,k}^{\rm Pre})
\end{align}
where $b^{\rm Int}_{i+1,j,k}=b(u^{\rm Int}_{i+1,j,k})$ and similar for $c_{i+1,j,k}$. However, the non-linearity of $b$ and $c$ implies that $b_{i+1,j,k}^G \not = b(u^G_{i+1,j,k})$.  To obtain a consistent value for $b_{i+1,j,k}^G$ we instead use
\begin{align}
& b_{i+1,j,k}^G \equiv b(u^G_{i+1,j,k})\\
& c_{i+1,j,k}^G \equiv c(u^G_{i+1,j,k}).
\end{align}
When calculating the differential operator for cells close to the boundary, the only changes from this choice are in the actual values we use for $b_{i+1,j,k} = b_{i+1,j,k}^G$, $c_{i+1,j,k} = c_{i+1,j,k}^G$, and $u_{i+1,j,k} = u_{i+1,j,k}^G$, but for the derivative of the operator Eq.~(\ref{dbeq}), we do pick up some new terms such as $b_{i+1,j,k}^G$ depends on $u_{i,j,k}$ through Eq.~(\ref{ueq}). The extra terms we need to add to Eq.~(\ref{dbeq}) are
\begin{align}
\frac{\partial L_{i,j,k}}{\partial b_{i+1,j,k}} \frac{\partial b_{i+1,j,k}}{\partial u_{i,j,k}} + \frac{\partial L_{i,j,k}}{\partial u_{i+1,j,k}} \frac{\partial u_{i+1,j,k}}{\partial u_{i,j,k}}
\end{align}
where
 \begin{align}
\frac{\partial L_{i,j,k}}{\partial b_{i+1,j,k}} \frac{\partial b_{i+1,j,k}}{\partial u_{i,j,k}} &=  c_{i+1,j,k}\frac{(u_{i+1,j,k}-u_{i,j,k})}{2h^2} \frac{m_{i+1,j,k}}{m_{i,j,k}}\\
\frac{\partial L_{i,j,k}}{\partial b_{i+1,j,k}} \frac{\partial b_{i+1,j,k}}{\partial u_{i,j,k}} &=  \frac{b_{i+1/2,j,k}}{h^2}\frac{m_{i+1,j,k}}{m_{i,j,k}}. 
\end{align}
We emphasise that these terms should only be included when $m_{i+1,j,k} < 0$, i.e. when the $(i+1)$'th cell is masked otherwise as $\frac{\partial u_{i+1,j,k}}{\partial u_{i,j,k}} \equiv 0$. The case where several neighbour cells are masked is covered by summing these expressions over all the masked cells.

\section{The equations and implementation details for specific models}
\label{section:specific_models}

In this section we describe the definition of the two models that we have implemented (the symmetron and $f(R)$ gravity), the dimensionless form of their equations, and details on the implementation.

\subsection{Symmetron model}

The symmetron model was originally proposed in \citet[][]{2010PhRvL.104w1301H} as a screening mechanism that allows a scalar field to mediate a long range ($\sim$ Mpc) force of gravitational strength in the cosmos while satisfying solar system tests of gravity. N-body simulations of this model have already been run, for instance, in \citet[][]{2012ApJ...748...61D} using a modified version of the code \tt{MLAPM} \citep[][]{Knebe01}. The effect of non-static terms in these simulations was studied by \citet[][]{2013PhRvL.110p1101L}.  The potential and conformal factor that defines the model are
\begin{align}
V(\phi) &= -\frac{1}{2}\mu^2\phi^2 + \frac{1}{4}\lambda\phi^4 \\
A(\phi) &= 1 + \frac{1}{2}\left(\frac{\phi}{M}\right)^2, 
\end{align}
where $\mu$ and $M$ are mass scales, and $\lambda$ is a dimensionless constant.

Taking Eq.~(\ref{canonical_equation}) into account, we find that the static equation of motion for the scalar field reads as
\be
\nabla^2 \phi = a^2 \left[\left(\frac{\rho}{M^2}-\mu^2\right) \phi + \lambda \phi^3 \right]. 
\ee
It is convenient to work with a dimensionless scalar field $\chi$, which we obtain by normalising $\phi$ with its vacuum expectation value, 
\be
\phi_0 = \frac{\mu}{\sqrt{\lambda}}.
\ee
We substitute the free parameters of the model $(\mu, \lambda, M)$ by the range of the field that corresponds to $\rho=0$,
\be
\lambda_0 = \frac{1}{\sqrt{2}\mu}, 
\ee
a dimensionless coupling constant,
\be
\beta = \frac{\phi_0 M_{\text{Pl}}}{M^2},
\ee
and the expansion factor for which the background density takes the value for which the symmetry is broken in the cosmological background
\be
a_{SSB}^3 = \frac{\rho_0}{\rho_{SSB}} = \frac{\rho_0}{\mu^2 M^2}. 
\ee
By substituting these definitions in the equation of motion we obtain
\be
\nabla^2 \chi = \frac{a^2}{2 \lambda_0^2} \left[ \left(\frac{a_{SSB}}{a}\right)^3\eta\chi - \chi + \chi^3 \right], 
\label{symmetron_dimensionles}
\ee
where $\eta$ is the matter density in terms of the mean density at any given redshift. 

The modified geodesics given by Eq.~(\ref{geodesics_general}) take the following form for this model:
\be
\ddot{\mathbf{x}} + 2 H \dot{\mathbf{x}} + \frac{1}{a^2} \nabla\Phi + \frac{1}{a^2} \frac{\phi}{M^2}\nabla\phi = 0, 
\ee
which can be written as
\be
\label{geo_code}
\ddot{\mathbf{x}} + 2 H \dot{\mathbf{x}} + 
    \frac{1}{a^2} \nabla\Phi + 
    \frac{6\Omega_m H_0^2}{a^2} \frac{(\beta\lambda_0)^2}{a_{SSB}^3} \chi\nabla\chi = 0,
\ee
when introducing the dimensionless scalar field $\chi$ and the parameters $(\lambda_0, \beta, a_{SSB})$.

The equation can be simplified further by using the super-comoving quantities defined by \citet[][]{1998MNRAS.297..467M}
\begin{align}
d \tau & = \frac{1}{a^2} dt\\
\tilde{\Phi} & = a^2 \Phi. 
\end{align}
A similar change in the scalar field
\be
\tilde{\chi} = a \chi
\label{change_chi_a_chi}
\ee
will also remove the explicit dependence with $a$ in the term related to the fifth force. In these variables, the equation becomes
\be
\frac{d^2\mathbf{x}}{d\tau^2} + \nabla\tilde{\Phi} + 6\Omega_m H_0^2 \frac{(\beta\lambda_0)^2}{a_{SSB}^3}\tilde{\chi}\nabla\tilde{\chi} = 0,
\label{symmetron_geodesics_tilde_chi}
\ee
which is the expression we use to evolve the positions of the particles in the N-body code.
This equation is solved by using the same leap-frog scheme as is included in the standard code.  The evolution scheme for the time step $n$ is given by the following equations:
\begin{align}
\label{leap_frog_1}
\mathbf{v}^{n+1/2} & = \mathbf{v}^n - \left[\nabla\phi^n\Delta + 6\Omega_m H_0^2 \frac{(\beta\lambda_0)^2}{a_{SSB}^3}\tilde{\chi}^{n}\nabla\tilde{\chi}^{n} \right] \tau/2\\
\mathbf{x}^{n+1} & = \mathbf{x}^n + \mathbf{v}^{n+1/2} \Delta \tau \\
\label{leap_frog_3}
\mathbf{v}^{n+1} & = \mathbf{v}^{n+1/2} - \left[\nabla\phi^{n+1} + 6\Omega_m H_0^2 \frac{(\beta\lambda_0)^2}{a_{SSB}^3}\tilde{\chi}^{n+1}\nabla\tilde{\chi}^{n+1} \right]\Delta \tau/2.
\end{align}
In the same way as in the standard code, the second evaluation of the velocities is made in the next time step to avoid calling both gravitational solvers twice in each time step.  The form of the evolution equation, Eq.~(\ref{symmetron_geodesics_tilde_chi}), is the same as for standard gravity, namely acceleration equals force given by the gradient of a potential. The main difference is that the force term is on average larger than for standard gravity. \tt{RAMSES} uses adaptive time-stepping to prevent particles moving to far each time step, so we expect the leap-frog scheme to work just as well for scalar tensor theories as it does for standard gravity. The time steps will in general be smaller to account for the stronger force.

\subsection{Hu-Sawicki f(R) model}
\label{section:hw}
As an example of applying the non-canonical equations, we have implemented the Hu-Sawicki $f(R)$ gravity model \citep[][]{2007PhRvD..76f4004H}. The model is originally defined in the Jordan frame through a modified Einstein-Hilbert term $R\to R+f(R)$ where $R$ is the Ricci scalar and $f$ is a free function.  The action that defines the model is
\be
S = \int \sqrt{-g} \left[ \frac{R+f(R)}{16\pi G} + \mathcal{L}_m \right] d^4 x, 
\ee
where $\mathcal{L}_m$ is the matter Lagrangian, and $f$ is chosen as
\be
f(R) = - m^2\frac{c_1(R/m^2)^n}{c_2(R/m^2)^n+1}, 
\ee
where $m^2 \equiv H_0^2\Omega_m$ and $c_1$, $c_2$ and $n$ are dimensionless model parameters. These three free parameters can be reduced to only two ($n$ and $f_{R0}$) by requiring the model to yield dark energy (here in the form of an effective cosmological constant). This requires
\be
\frac{c_1}{c_2} = \frac{6\Omega_\Lambda}{\Omega_m}. 
\ee
Instead of using $c_1$ (or $c_2$) as our second free parameter, it is convenient to use
\be
f_{R0} = -n\frac{c_1}{c_2^2}\left(\frac{\Omega_m}{3(\Omega_m+4\Omega_\Lambda)}\right)^{n+1}, 
\ee
which is related to the range of fifth force in the cosmological background today via 
\begin{align}
\lambda_\phi^0 = 3 \sqrt{\frac{(n+1)}{\Omega_m+4\Omega_\Lambda}}\sqrt{\frac{|f_{R0}|}{10^{-6}}}~~~ \mbox{Mpc}/h.
\end{align}
The $f(R)$ models can be transformed into a scalar-tensor theory in the form of the action given by Eq.~(\ref{action_scalar_tensor}) through a Weyl transformation
\be
\tilde{g}_{\mu\nu} = A^2(\phi)g_{\mu\nu}
\ee
where \cite[][]{brax}
\begin{align}
f_R = e^{-\frac{2\beta\phi}{M_{\rm Pl}}}-1 \simeq -\frac{2\beta\phi}{M_{\rm Pl}}
\end{align}
with $\beta = \frac{1}{\sqrt{6}}$. This equation defines $R(\phi)$, which can be used to get the potential $V(\phi)$ that is given by
\begin{align}
V(\phi) &= \frac{M_{\rm Pl}^2(f_RR -f)}{2(1+f_R)^2}. 
\end{align}
The resulting equation of motion for the scalar field $f_R$ in the static limit is
\begin{align}
\nabla^2f_R = &-\frac{1}{a}\Omega_m H_0^2\left(\eta - 1\right)  + a^2\Omega_m H_0^2 \times \nonumber\\
&\times \left[\left(1+4\frac{\Omega_\Lambda}{\Omega_m}\right)\left(\frac{f_{R0}}{f_R}\right)^{\frac{1}{n+1}}  - \left(a^{-3} + 4\frac{\Omega_\Lambda}{\Omega_m}\right)\right], 
\end{align}
where $f_{R0}$ is the value that corresponds to the minimum for the background density today and can be written as in Eq.~67.  Since we work in the Einstein frame, the equation for the metric perturbations will be the standard Poisson's equation. This is a different implementation\footnote{By introducing the (Jordan-frame) potential $\Phi_J = \Phi - \frac{f_R}{2}$ ($\tilde{\Phi}_J = \tilde{\Phi} + \frac{1}{2}e^u$) one can transform the equations to that of \cite{2008PhRvD..78l3523O} and \citet{2012JCAP...01..051L}. Poisson's equation for $\Phi_J$ follows from simply adding Poisson's equation for $\Phi_N$ and the Klein-Gordon equation for $-f_R/2$.} (though mathematically equivalent) from what is done in other codes that have implemented this model \citep{{2008PhRvD..78l3523O}, {2012JCAP...01..051L},{2013arXiv1305.2418P}}. 
As noted in \cite{2008PhRvD..78l3523O}, the scalar field equation of motion can be written in a more numerically stable form by making a field redefinition:
\be
f_R = -a^{-2}e^u.
\ee
The equation of motion in its non-canonical form is then
\begin{align}
&\nabla\cdot\left(b(u)\nabla u\right) =\Omega_m a H_0^2(\tilde{\rho}-1)\nonumber\\
&- \Omega_m a^4 H_0^2\left(1 + 4\frac{\Omega_\Lambda}{\Omega_m}\right)(|f_{R0}|a^2)^{\frac{1}{n+1}}e^{-\frac{u}{n+1}}\nonumber\\
&+\Omega_m a H_0^2\left(1 + 4a^3\frac{\Omega_\Lambda}{\Omega_m}\right)
\end{align}
where $b(u) = e^u$.  The discretization of this equation was implemented as described in Sect. \ref{discretiz_non_canonical}.  The geodesic equation reads as
\begin{align}
\frac{d^2{\bf x}}{d\tau^2} + \nabla\tilde{\Phi} + \frac{1}{2}e^u\nabla u = 0.
\end{align}
The evolution scheme used for the geodesics is equivalent to the one used for the symmetron model and in the standard code (Eqs. (\ref{leap_frog_1}) to (\ref{leap_frog_3})).

\begin{figure*}[t!]
\begin{minipage}[t]{0.49\textwidth}
\includegraphics[width=1.0\textwidth]{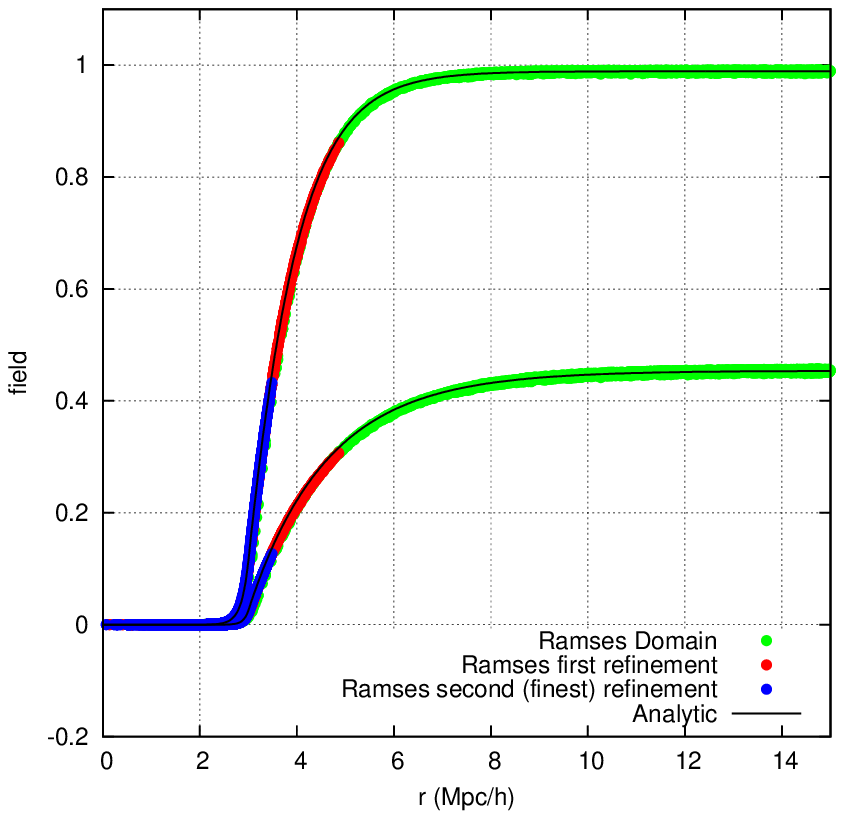}
\end{minipage}
\hfill{}
\begin{minipage}[t]{0.49\textwidth}
\includegraphics[width=1.0\textwidth]{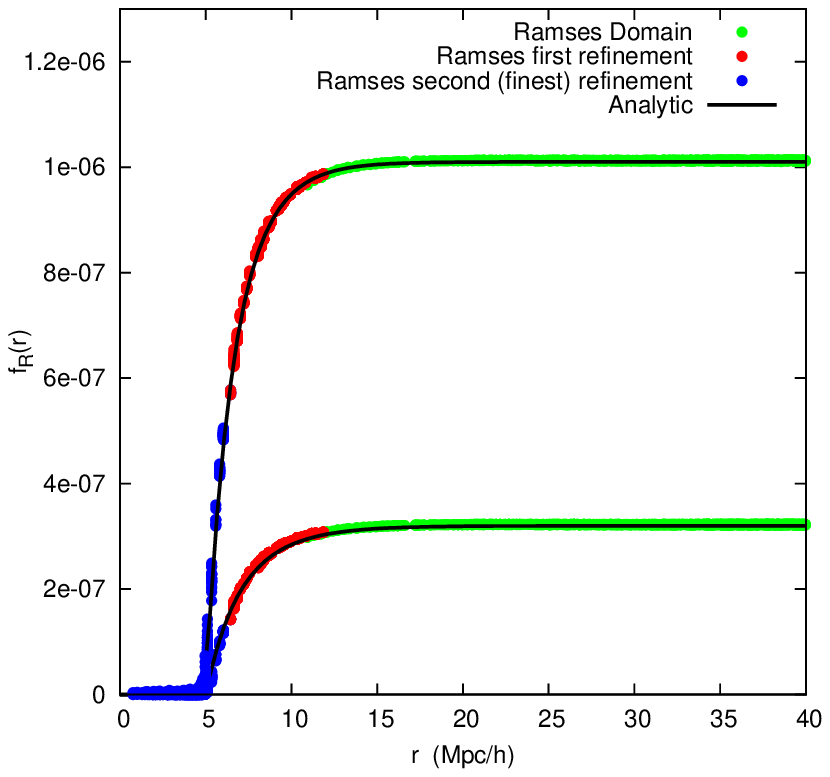}
\end{minipage}
\caption{Left:  comparison between analytical and numerical solutions of the symmetron model for two different redshifts.  The density distribution is given by a sphere of constant density.  Different colours correspond to different levels of refinement.  The thin line is the analytic solution for redshift $z=0$ (upper line) and $z=1$ (lower line).  Right:  same for the $f(R)$ solver.  See text for details.}
\label{fig:static_tests}
\end{figure*}

\section{Code tests}
\label{section:tests}

In this section we show the results of the tests that were performed to our implementation of the scalar field solvers.

\subsection{Potential solver}

To measure the quality of the new solver, we compared results with solutions obtained for a sphere of uniform density located in the centre of the box.  Confident solutions to compare with can be obtained by writing the equation in spherical coordinates and solving the resulting 1D equation with standard packages such as Mathematica.  

The density corresponds to a sphere of radius $R$ of constant density embedded in a uniform background:
\be
\rho(r) = \begin{cases}
  \rho_\text{in} = (1+\delta) \frac{\bar{\rho}}{1+\frac{4\pi}{3}\delta\left(\frac{R}{B}\right)^3} & \text{$r < R$}, \\
  \rho_\text{out} = \frac{\bar{\rho}}{1+\frac{4\pi}{3}\delta\left(\frac{R}{B}\right)^3} & \text{$r > R$},
  \end{cases}
\ee
where 
\be
\delta=\frac{\rho_\text{in}-\rho_\text{out}}{\rho_\text{out}}
\ee
characterises the jump in the density, $\bar{\rho}$ is the mean density in the box, $R$ the radius of the sphere, and $B$ the size of the box.  The density is provided to the code through a distribution of particles.  The density estimation (CIC) and refinement criteria are the same as those used for the cosmological simulations we performed. The value of $\delta$ chosen for the test is 5000. The radius and box size for the symmetron test was taken to be $R=3~\text{Mpc}/h$ and $B = 40$ Mpc/h, respectively. For the $f(R)$ test we used $R=5$ Mpc/h and $B = 100$ Mpc/h. For both tests we used $128^3$ particles and a domain grid with 128 nodes per dimension.  To test that the treatment of the boundary of the refinement is correct, we included two levels of refinement.  

In the symmetron case, the test was made at redshift $z=0$ and $z=1$, and the models parameters were defined as $\lambda_0=1~ \text{Mpc}/h$ and $a_{SSB}=0.6$.  This gives us the possibility to test our solver in a situation in which the redshift is higher than the symmetry breaking redshift, but the density outside the sphere is low enough for the field to have a non-zero expectation value.  The model parameters for the test of the $f(R)$ code are $n=1$ and $|f_{R0}|=10^{-6}$. Figure \ref{fig:static_tests} shows the result of the tests for the symmetron (left) and $f(R)$ (right) codes.  The continuous line is the 1D solution, and the points are the solution on the grid that was obtained using the new solvers.  The different colours depict the different refinement levels. The test was performed using both the serial version and the parallel version running with eight processes.  Both versions gave the same results, showing that the parallel version of the solver works properly.

\subsection{Time evolution}

\begin{figure*}[t!]
\begin{minipage}[t]{0.49\textwidth}
\includegraphics[width=1.0\textwidth]{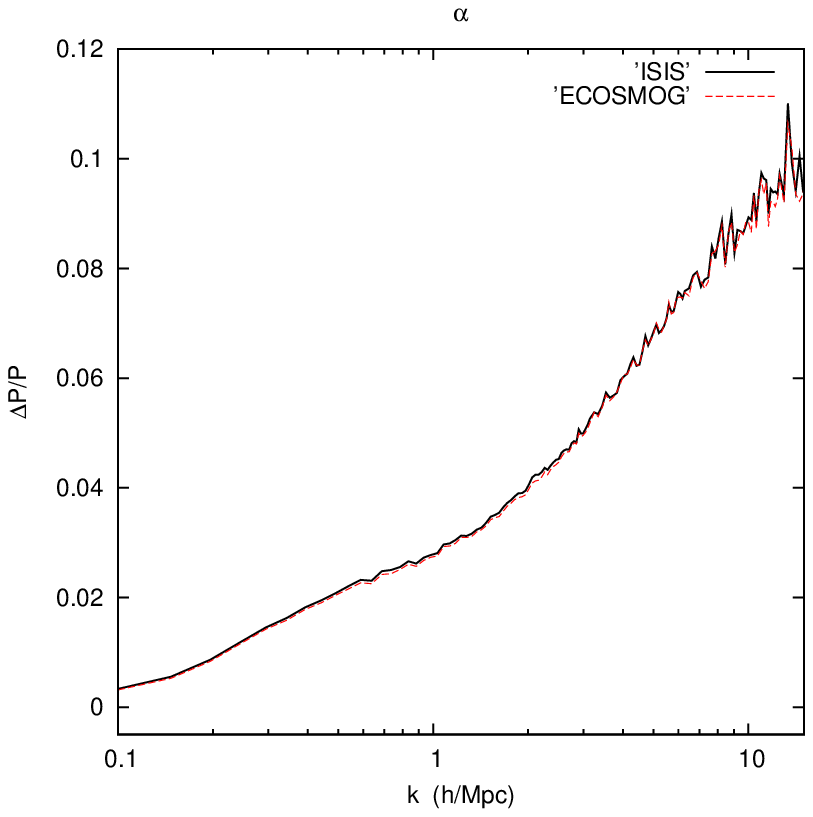}
\end{minipage}
\hfill{}
\begin{minipage}[t]{0.49\textwidth}
\includegraphics[width=1.0\textwidth]{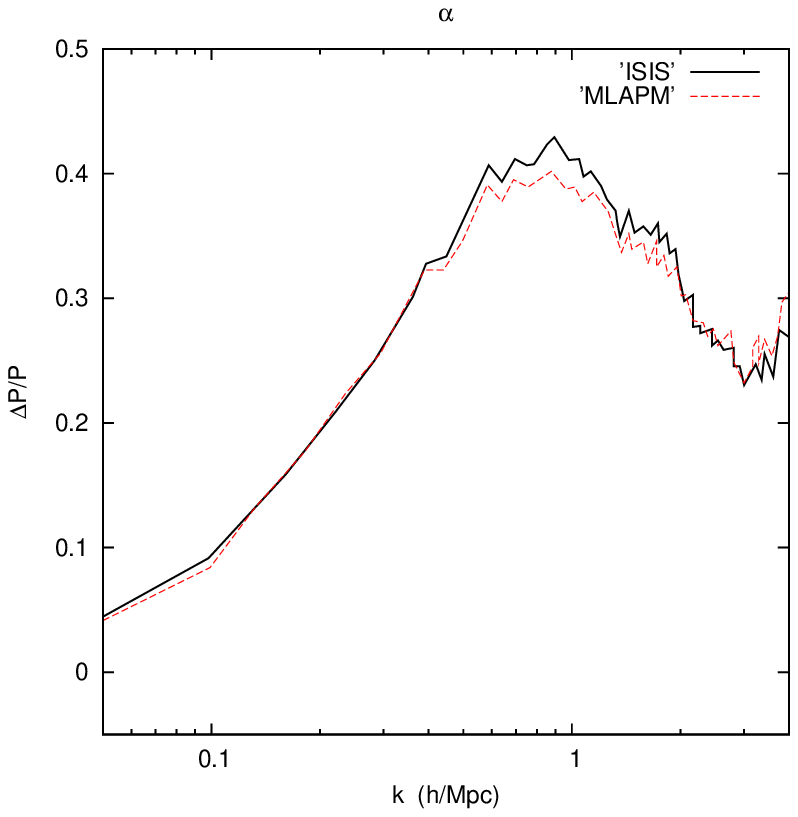}
\end{minipage}
\caption{Relative difference in the symmetron matter power spectrum with respect to $\Lambda$CDM for our code (red) and a similar implementation of the code \tt{ECSOMOG}. Right:  same comparison with the code \tt{MLAPM}.  See text for details of the models and simulations parameters employed.}
\label{fig:comparison_p_of_k_symmetron}
\end{figure*}

To test the time evolution of the new code, we ran cosmological simulations and compared the final matter power spectrum at redshift $z=0$ with results that were obtained with similar codes or taken from the literature.  In the symmetron case, we ran simulations with three different codes:  our new code that we wanted to test, a modification of the code \tt{MLAPM} \citep{mlapmcode}, and the code \tt{ECOSMOG} \citep[][]{2012JCAP...01..051L}.  The initial conditions for the comparison with the code \tt{ECOSMOG} were the same as were used in the simulations presented in  \citet[][]{2012JCAP...01..051L} and were constructed using a box of 128 Mpc/$h$ and $256^3$ particles.  The symmetron parameters of this particular simulation are $(a_{\text{SSB}}, \lambda_0, \beta) = (0.5,1~\text{Mpc}/h, 1)$, while the background cosmology is given by $(\Omega_m, \Omega_{\Lambda}, H_0)=(0.267,0.733,71.9 ~ \text{km/sec/Mpc})$.  Both simulations were run using the same random seed for the realisation of the initial density field, which implies that the differences that we find between different runs can only be attributed to differences between the codes and not to cosmic variance. In an effort to isolate differences that could exist in the modified gravity part of the codes from the $\Lambda$CDM part, we also ran simulations using standard gravity with both codes using the same initial conditions and the same background cosmology.  The comparison is then made on the difference between the standard and modified gravity codes.  Details on the particular implementation of \tt{ECOSMOG} used for the test can be found in \citet[][]{2012JCAP...01..051L}. In brief, the code includes a generalised description for scalar fields, for which the symmetron model is a limit case.  The left-hand panel of figure \ref{fig:comparison_p_of_k_symmetron} shows the relative difference between the power spectrum at redshift $z=0$, which was obtained from the symmetron and $\Lambda$CDM simulations with both codes. The differences between the two codes are below 0.2\% for all $k< 10~h/$Mpc.

The initial conditions for the comparison with the \tt{MLAPM} code were generated for a box of 64 Mpc/h and with $128^3$ particles. As in the previous test, the simulations with \tt{ISIS} and \tt{MLAPM} were run with exactly the same initial conditions and background cosmology, but in this case we used the symmetron parameters  $(a_{\text{SSB}}, \lambda_0, \beta) = (0.33,1~\text{Mpc}/h, 1)$, which means that the symmetry is broken at earlier times in the background, and so the effects of the fifth force are enhanced.  As described above, we also used $\Lambda$CDM simulations that were run with the \tt{ISIS} and \tt{MLAPM} codes as reference. The right-hand panel of figure \ref{fig:comparison_p_of_k_symmetron} shows the outcome of the test.  A different treatment of the refinement structure and in time stepping increases the differences between both codes, but the difference is still below 1-2\% for all $k< 4~h/$Mpc.

\begin{figure}[t!]
\includegraphics[width=1\columnwidth]{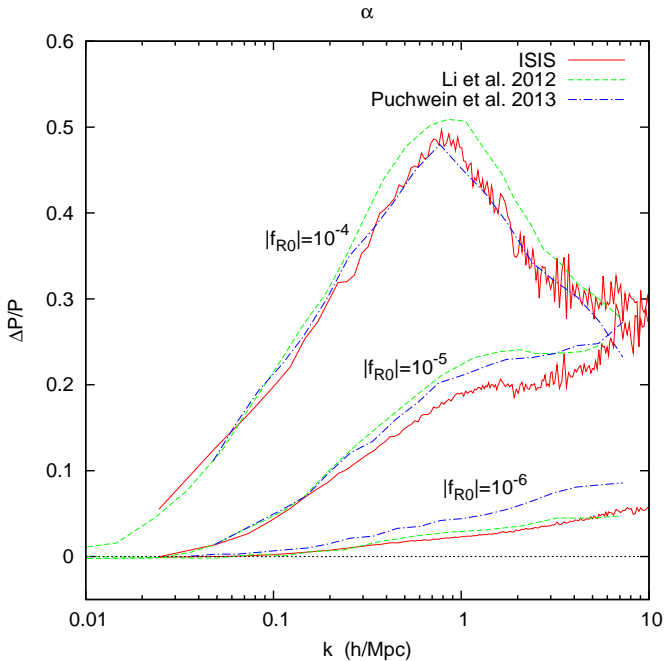}
\caption{Relative difference in the $f(R)$ power spectrum with respect to $\Lambda$CDM for our code (red) and similar implementations published in \citet[][]{2012JCAP...01..051L} and \citet[][]{2013arXiv1305.2418P}.  Different sets of curves correspond to different values of $f_{R0}$.}
\label{fig:comparison_p_of_k_fofr}
\end{figure}

To test the time evolution of the $f(R)$ code, we compared the results of our simulations with others taken from the literature.  The codes used in the works that we take as reference are \tt{ECOSMOG} \citep[][]{2012JCAP...01..051L} and \tt{Gadget} \citep[][]{2013arXiv1305.2418P}.  Both codes include the same $f(R)$ model that we included in our code, but their implementation was made in the Jordan frame instead of the Einstein frame that we decided to use.  We ran three simulations using $f_{R0}=10^{-6}, 10^{-5}$, and $10^{-4}$, which goes from an almost Newtonian limit model to a model that includes strong effects of the fifth force.  Figure \ref{fig:comparison_p_of_k_fofr} shows the power spectrum of the three $f(R)$ simulations with respect to a $\Lambda$CDM run that was made for comparison using the same initial conditions.  The three simulations were run using different initial seeds for the random number generator used to calculate the initial conditions, and thus the initial phases are different. Cosmic variance should be taken into account when comparing the curves that correspond to each code.\footnote{Since the figure shows the relative difference with respect to $\Lambda$CDM, the cosmic variance will not appear on large scales, but rather on intermediate and small scales.}  The simulations of the three codes still agree remarkably well on scales $0.01~h/\text{Mpc}\lesssim k\lesssim 1~h/$Mpc.

\section{Application:  shape of clusters as a test for modified gravity}
\label{section:application}

As an application of our new code, we test the impact that the scalar field fifth force has on the haloes shape of groups and clusters of galaxies.  \citet[][LLM13 from now on]{2013PhRvL.110o1104L} proposed to study this observable by using static calculations of the total gravitational potential.  The general result is that the presence of a scalar field increases the ellipticity of the total gravitational force distribution, hence of the x-ray emitting gas residing in the clusters.  Given the observational constraints that exist on the ellipticity of the x-ray component of clusters, LLM13 gave constraints on the model parameters $\beta$ and $\lambda_0$ (coupling constant and range) for chameleon \citep{cham} and symmetron models.

A back-of-the-envelope calculation can help for understanding this result.  The calculation consists in applying a perturbation to a spherical object and analysing the shape of the associated gravitational potential.  One could, for instance, propose the following perturbed density-potential pair (we focus on the region close to $\theta = 0$):
\begin{align*}
  \left\{\begin{aligned}
  & \rho_{DM}(r,\theta) = \rho_0(r) \left( 1 + \epsilon_{\rho}\cos^2\theta\right)\\
  & \phi(r,\theta) = \phi_0(r) \left( 1 + \epsilon_{\phi}\cos^2\theta\right)\\
  \end{aligned}\right.
\end{align*}
where $\rho_0$ and $\phi_0$ is the pair associated to the unperturbed spherical system, and $\epsilon_{\rho}$ and $\epsilon_{\phi}$ are small numbers.  By substituting this pair in Poisson's equation
\be
\nabla^2\phi = 4\pi G \rho
\ee
and keeping only first-order terms in the perturbations, one obtains
\be
\epsilon_{\phi} =  \frac{4\pi G \rho_0 r^2 \sin\theta}{4\pi G \rho_0 r^2\sin\theta-\frac{2\phi_0\cos(2\theta)}{r^2\cos(\theta)^2}} \epsilon_{\rho}, 
\ee
which in the limit $\theta\rightarrow 0$ (i.e. maximizing the perturbation) gives $\epsilon_{\phi} \rightarrow 0$.  In other words, in the Newtonian case, the gravitational potential is insensitive to perturbations in the shape of the density up to first order, which is a well known result (see for instance results from simulations in \citealt[][]{2011ApJ...734...93L}).

By repeating the same analysis for instance with the equation for the chameleon field \citep[][]{cham}
\be
\nabla^2\phi_c  = \frac{\beta}{M_{pl}}\rho - \frac{M^5}{\phi^2}, 
\ee
we obtain 
\be
\epsilon_{\phi} = \frac{\epsilon_{\rho}}{\left[ 1 - \left( \frac{3 M^5}{\phi_0^2} + \frac{2\phi_0\cos(2\theta)}{\sin\theta\cos(\theta)^2r^2}\right) \frac{M_{pl}}{\beta \rho_0} \right] }.
\ee
If we now take the limit $\beta\rightarrow \infty$ we find $\epsilon_{\phi} \rightarrow \epsilon_{\rho}$, so in the largely coupled limit, the modified model is sensitive to changes in the shape of the density up to first order. An alternative way of looking at the same problem is the following.  In the chameleon case, when $\beta$ or $\rho_0$ is large enough, the field will be forced to stay close to the minimum of its effective potential $\phi_{\rm min} = (\beta\rho/M^5M_{\rm Pl})^{-1/2}$, which is uniquely determined by the local matter density. 
This implies that in the limit $\beta\rho \to \infty$, the iso-contours of the scalar field will be completely aligned with the iso-contours of the matter density, no matter how this density is distributed. 

By performing numerical calculations for a fixed density distribution, LLM13 proved that this result can be also extended to symmetron models.  However, their result is incomplete in the sense that time evolution was not taken into account in the calculations; in other words, the dark matter distribution was fixed and the back reaction of the difference found in the potential into the shape of the dark matter halo itself was not taken into account.  We give here the next natural step towards a fully realistic analysis, by including the time evolution in the dark matter component.  Since in the present version of the \texttt{ISIS} code, baryonic physics is not included, for the moment we can only give constraints on the shape of the dark matter haloes.  We remind the reader that the shape of the baryonic component is still not fully understood within the context of standard gravity and that the so-called over-cooling problem is still not completely settled \citep[see e.g.][]{2011ApJ...734...93L}.  The impact of modified gravity on the shape of the gas distribution is left for future work.

To properly interpret the results that will be presented below, it is important to understand the relation between the models studied in this paper and those presented in LLM13.  The symmetron model employed is the same in both papers.  However, here we substituted the chameleon studied in LLM13 with the Hu-Sawicki $f(R)$ model \citep[][]{2007PhRvD..76f4004H} described in Sect. \ref{section:hw}. Both of the two models have an equation of the following form in a non-cosmological setting:
\be
\nabla^2 \phi = - \frac{M^{4+n}}{\phi^{1+n}} + \frac{\beta}{M_\text{pl}}\rho.  
\ee
The mapping between the Hu-Sawicki $f(R)$ model we have simulated and the chameleon model is given by $n=-1/2$, and $\beta= 1/\sqrt{6}$.  The constant $M$ can be written in terms of the range of the field in the cosmological background today via
\be
\lambda^0_\phi = \left(\frac{2M^7}{27\beta^3\Omega_mH_0^6M_{\rm Pl}^5}\right)^{\frac{1}{2}}.
\ee

\subsection{Simulations}

The data to be used for the analysis was obtained from a set of simulations that we ran with both standard gravity and the two models presented in Sect. \ref{section:specific_models}.  Table \ref{tab:model_parameters} summarises the model parameters.  The initial conditions were generated assuming that both scalar field models give fully screened solutions at high redshift, and thus the Zeldovich approximation is also valid in the modified models.  In practice, we generated only one set of initial conditions with the package Cosmics \citep[][]{1995astro.ph..6070B}.  We used a box size of $256$ Mpc/h and $512^3$ particles.  The background cosmology is also the same for all the simulations and is defined as a flat $\Lambda$CDM model given by $(\Omega_m, \Omega_{\Lambda}, H_0) = (0.267,0.733,71.9 ~ \text{km/sec/Mpc})$.

\begin{table}
  \begin{tabular}{lrrr}
    Model  & $\lambda_0$ & $z_{SSB}$ & $\beta$  \\
    \hline 
    Symm A & 1 & 1 & 1 \\
    Symm B & 1 & 2 & 1 \\
    Symm C & 1 & 1 & 2 \\
    Symm D & 1 & 3 & 1 \\
  \end{tabular}
  \begin{tabular}{lrrr}
    Model  & $n$ & $|f_{R0}|$ & $\lambda_0$ \\
    \hline 
    fofr4 & 1 & $10^{-4}$ & 23.7 \\
    fofr5 & 1 & $10^{-5}$ & 7.5 \\
    fofr6 & 1 & $10^{-6}$ & 2.4
  \end{tabular}
  \caption{Model parameters for the symmetron (left) and $f(R)$ (right) runs.  The range in the $f(R)$ model is derived from the value of $f_{R0}$ and given only to have a reference point to compare both families of models.  The range is given in Mpc/h in both set of models.}
  \label{tab:model_parameters}
\end{table}

\subsection{Analysis}

The aim of the analysis is to measure the shape of the dark matter haloes present in the simulations and to make the comparison between $\Lambda$CDM haloes and those that were formed under the influence of the modified gravity models.  The haloes were identified using the code \tt{Rockstar} \citep{2011arXiv1110.4372B}.  Since gravitational lensing observations measure the distribution of the total mass of the clusters, we did the analysis using all the mass included inside the virial radius.  We used a lower cut off of $10^{13} M_{\odot}$, which give haloes with more that $\sim 10^3$ particles.  The upper cut-off in mass is given naturally by the size of the box. In other words, we present results that are applicable to groups and clusters of galaxies.  We concentrate all our analysis at redshift $z=0$.

To avoid contaminating the results with unrelaxed clusters, we defined the virialisation state by taking into account the virial theorem, 
\be
\frac{1}{2}\frac{d^2I}{dt^2} = 2 T + W - E_s, 
\ee
where
\be
T = \frac{1}{2}\sum_p m_p v^2_p
\ee
is the total kinetic energy, 
\be
W = -\frac{1}{2}\sum_{p,q} \frac{m_p m_q}{|\mathbf{r}_p - \mathbf{r}_q|}
\label{def_w}
\ee
is the total potential energy, and $E_s$ is a surface pressure term that takes the effects of the environment  into account over the dynamical state of the haloes  \citep[e.g.][]{2006ApJ...646..815S}.  It is customary to define the virial ratio
\be
\eta = \frac{2T-E_s}{W} + 1.  
\ee
Virialized objects are defined as having $d^2I/dt^2\approx 0$, and thus $\eta\sim 0$.  However, one must consider that the value for $\eta$ is not exactly zero for virialized objects, but oscillates around that value. We define the virialized sample as composed of those haloes for which $|\eta|$ is below a certain threshold, which we choose as 0.2.  The definition given by Eq.~(\ref{def_w}) is only a lower bound of the true potential energy, which is likely to have higher values owing to the presence of the environment in which each dark matter halo is immersed.  Furthermore, the condition $\eta < 0.2$ that we use to define the virialized sample is consistent with the literature \citep[e.g.][]{2006ApJ...646..815S}, but ultimately arbitrary.  In this sense, the criteria that we employ to define the relaxed sample must be taken as a rough diagnostic rather than a definitive choise.

In the modified gravity case, the definition of the potential energy must be modified to take into account the energy of the scalar field.  In this case, the superposition principle is not valid anymore, so that the virialisation state cannot be calculated by using previous expressions.  A detailed study of this issue applied to the same data presented here is shown in a companion paper \citep[][]{2013arXiv1307.6994G}.  In the present work, we use a simplified analysis that consists in calculating the distribution of the $\eta$ parameter by using standard gravity and shifting it in such way that its maximum is centred on zero.  Because the models that we are treating include a mass-dependent screening, we calculated these distributions not for the complete sample, but for subsamples defined by four uniform logarithmic mass bins ranging from $10^{13}$ to $10^{15} M_{\odot}$.  

\begin{figure*}[t]
\begin{minipage}[t]{0.49\textwidth}
\includegraphics[width=1\textwidth]{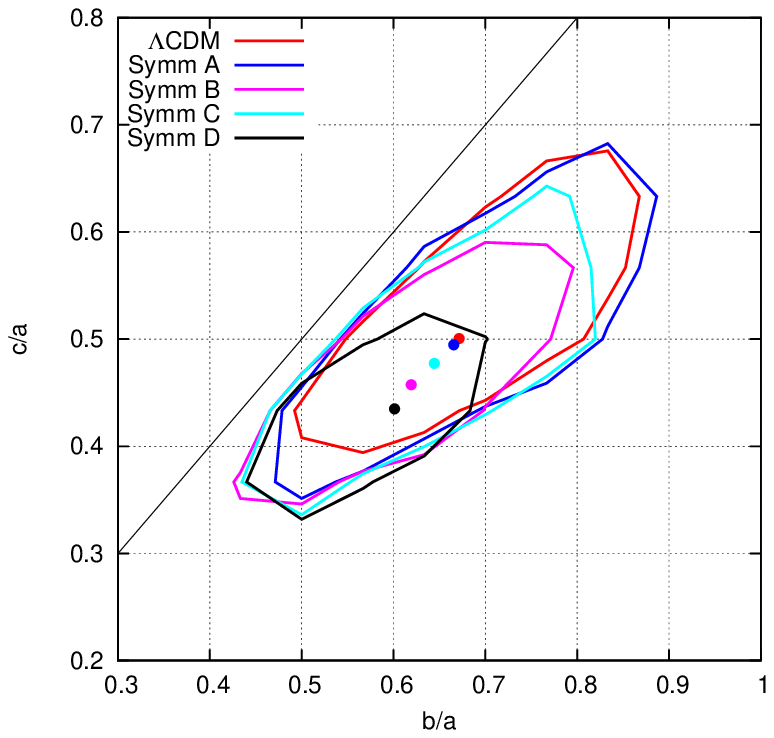}
\end{minipage}
\hfill{}
\begin{minipage}[t]{0.49\textwidth}
\includegraphics[width=1\textwidth]{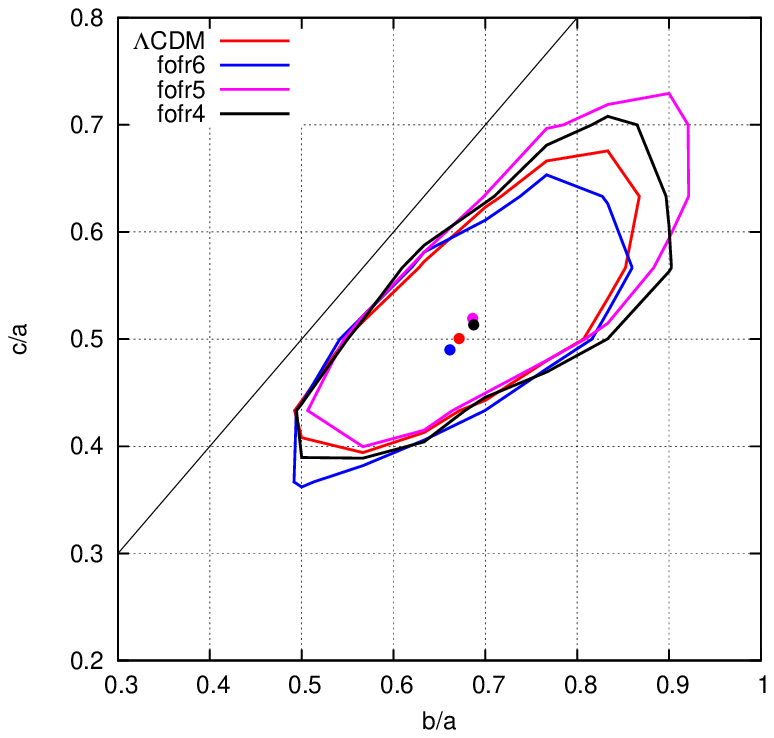}
\end{minipage}
\caption{Contours of the distribution of ratios between semi-axis for the symmetron (left) and $f(R)$ (right) simulations.  Both panels also include the $\Lambda$CDM simulation for comparison.  The points are the expected values of the distributions.  The size of the error bars of these are comparable to the size of the points and are not included in the plot. }
\label{fig:ca_of_ba}
\end{figure*}

Once we have specified the sample of halos, we then measure the shape of their density distribution, which can be defined starting from
\be
M_{ij} = \int \rho x_i x_j d^3x.
\label{definition_shape}
\ee
 The iso-density surfaces can be approximated by ellipsoids described by the radial ellipsoidal coordinate 
\be
k = \sqrt{x^2+\frac{y^2}{q^2} + \frac{z^2}{s^2}}
\label{def_k}
\ee
with axial ratios
\be
 q^2=\frac{M_{xx}}{M_{zz}}  \text{ and }   s^2=\frac{M_{yy}}{M_{zz}}, 
\ee
where $M_{xx}$, $M_{yy}$ and $M_{zz}$ are the eigenvalues of $M_{ij}$.  The integral in Eq.~(\ref{definition_shape}) is computed by summing over the particles,
\be
M_{i,j} = \sum_l m_l x_i x_j,
\ee
up to the virial radius $R_{200}$, where $m_l$ is the mass of the particle $l$, which is the same for all the particles in our simulations.  The shape of the regions in which the integral is calculated for each halo is a triaxial ellipsoid that was determined iteratively as in \citet[][]{1991ApJ...378..496D}.

\subsection{Results}

Figure \ref{fig:ca_of_ba} shows the distributions of the ratios between semi-axis $q=b/a$ and $s=c/a$ for the whole sample found in each simulation, with expectation values of the distributions.  The error of this quantities is comparable to the size of the dots and are thus not shown.  The results from the symmetron simulations are shown in the left-hand panel and can be compared directly with those presented in LLM13.  In there, the dependency of the shape of the iso-surfaces of the scalar field was shown as a function of $z_{SSB}$, $\lambda_0$, and $\beta$.  The general result is that these iso-surfaces move away from the iso-potentials when decreasing both $z_{SSB}$ or $\lambda_0$.  The dependence on $\beta$ is the opposite: the higher the coupling, the more elliptical these iso-surfaces are.  The simulations presented here show the expected dependence with $\beta$: the Symm C simulation have lower ratios than the Symm A and $\Lambda$CDM simulations.  However, in the case of $z_{SSB}$, we find a behaviour that is opposite to what we expected:  the higher $z_{SSB}$, the lower the ratios found in the simulation, thus the more elongated the haloes.  The result is a direct consequence of taking the time evolution of the haloes into account.  While according to LLM13, a low value for $z_{SSB}$ increases the ellipticity of the iso-surfaces of the scalar field, the objects in these models are only influenced during a short time by the fifth force.  When considering the time evolution, one takes not only the intrinsic distribution of the scalar field into account , but also its history and cumulative effects, which turn the results upside down.

\begin{figure*}
\begin{minipage}[t]{0.49\textwidth}
\includegraphics[width=1\textwidth]{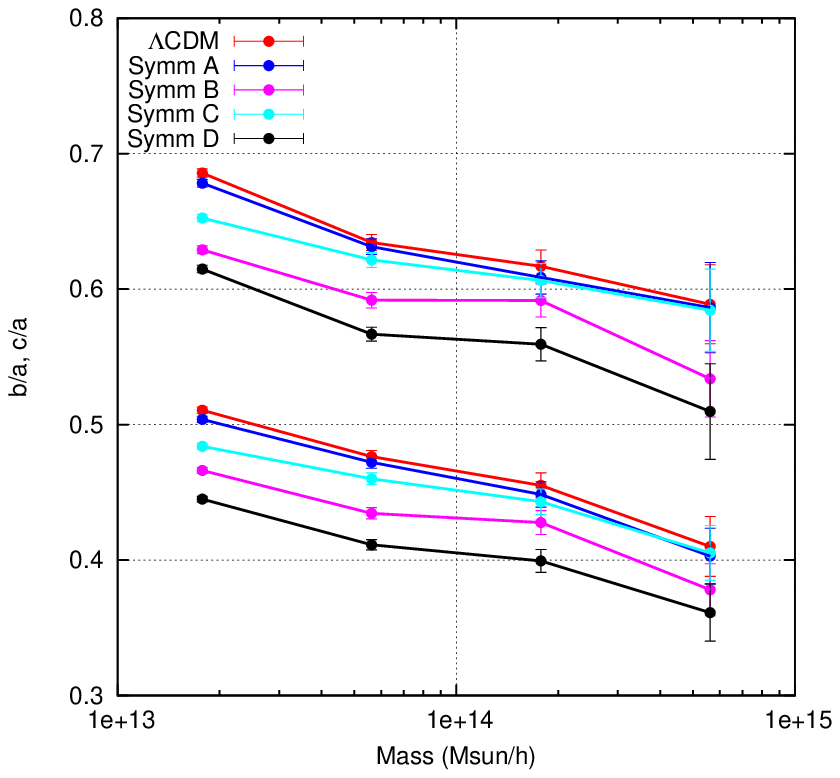}
\end{minipage}
\hfill{}
\begin{minipage}[t]{0.49\textwidth}
\includegraphics[width=1\textwidth]{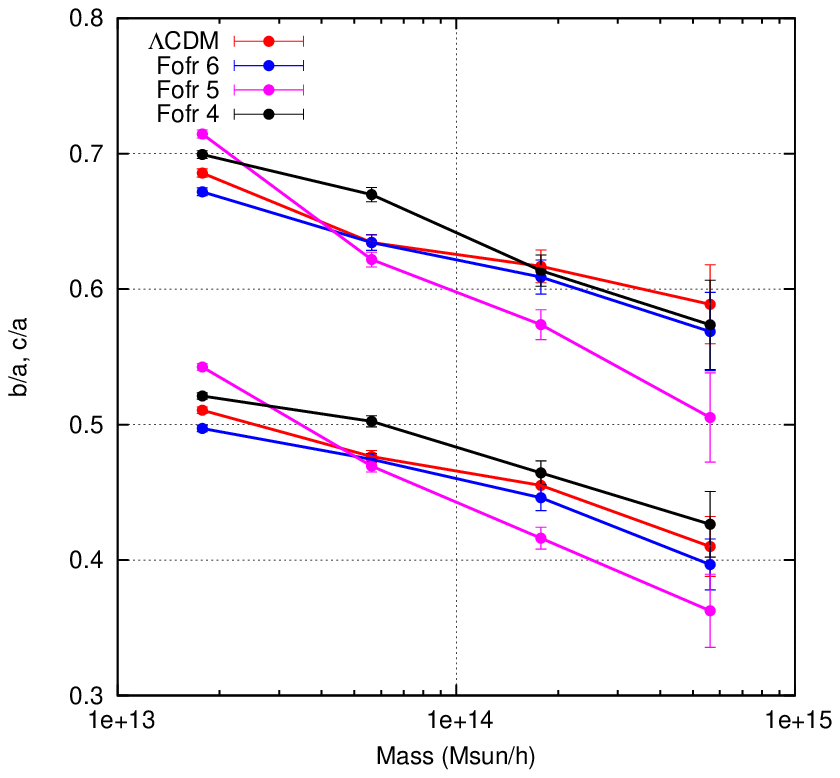}
\end{minipage}
\caption{Axial ratios $b/a$ (upper group of lines) and $c/a$ (lower group of lines) obtained from the symmetron (left) and $f(R)$ (right) simulations as a function of halo mass.  Both panels also include the results from the $\Lambda$CDM simulation for comparison.}
\label{fig:ratios_of_mass}
\end{figure*}

In the $f(R)$ case (right panel of figure \ref{fig:ca_of_ba}), there is not a clear trend of the shapes with respect to the free parameter $f_{R0}$.  The expectation value of the distributions moves slightly towards more elliptical objects in the fofr6 runs and towards more spherical objects in the other two.  An analysis of the mass dependence of this distributions will help for interpreting this results.

The simulations give us the chance to extend the results presented in LLM13 by taking the mass dependence of the signal into account.  We show this dependency in figure \ref{fig:ratios_of_mass}.  We find not only that the changes in shape are not given equally at all masses, but also that haloes of different mass are affected in a different way.  In addition to this, we find this dependence is not equal for all the models, but that each model has a different behaviour.  In the symmetron case, we find, within the region of the parameter space tested with this simulations, that increasing $z_{SSB}$ while fixing $\lambda_0$ and $\beta$ (see simulations Symm A, B, and D) increases the ellipticity almost evenly in the whole mass range studied.  The change in the shapes when modifying $\beta$ (see simulations Symm A and C) occurs only a low masses.

The dependence of the shapes on mass found in the $f(R)$ simulations is shown in the right-hand panel of figure \ref{fig:ratios_of_mass}.  The fofr6 model (for which we found an increase in the ellipticity in the previous analysis) shows that the increase in the total distribution is given only by the low mass haloes.  The high-mass haloes are mostly screened, so there is not fifth force capable of changing the shapes.  In the low-mass haloes, even if the force can be screened in their centre, it is still active in the outer regions, so it is possible to find differences with respect to $\Lambda$CDM (see \citet[][]{2013arXiv1307.6994G} for a study of the force distribution in the haloes of this simulations).  In the fofr5 case, we find that the near absence of deviations from $\Lambda$CDM shown in figure \ref{fig:ca_of_ba} is actually produced by a compensation given by a large deviation towards more elliptical haloes at high-mass and a correction towards more spherical haloes in the low-mass end.  While the deviation in the high-mass end is much larger, there are many more low-mass haloes, so that the effect of all of them together is enough to change the expectation of the whole sample towards more spherical haloes.  Finally, the fofr4 case seems to contradict the results presented in LLM13. The haloes in this model are actually slightly more spherical than in the $\Lambda$CDM case in the low-mass half of the sample.  By extrapolating results from LLM13 towards a model with a range $\lambda_0=23.7$ and $n=-1/2$ (according to the notation in LLM13), one could expect that the shapes should be similar to $\Lambda$CDM. To explain why fofr4 haloes are more spherical, one should enrich the analysis with more information than just the distribution of forces in a static case.  A possible mechanism for making the axial ratios even closer to one than in standard gravity could be based on the fact that haloes are not formed by spherical collapse, but in a hierarchical way.  To take the formation history of the haloes into account could help in understanding this result.  For instance, it was proven by using N-body simulations that modified gravity models have the property of increasing the collisional velocity of dark matter haloes \citep[][]{2009ApJ...695L.145L, 2012ApJ...747...45L}.  The model fofr4 is a model in which the fifth force acquires extreme values, and thus, the collisional velocity in mayor mergers should be greater than in $\Lambda$CDM, which should change the way in which haloes move towards equilibrium, making them more spherical.

\section{Conclusions and discussion}
\label{section:conclusions}

Several extensions of the standard cosmological model include scalar fields as new degrees of freedom in the underlying gravitational theory, which can be, for instance, in the form of scalar, vector, or tensor fields.  In general, these new degrees of freedom interact with matter and, in particular, with the standard model fields. Since deviations from Einstein gravity have neither been observed nor measured up to now in the solar system, these interacting scalar field theories must include screening mechanisms intended to hide the scalar field below observational limits within the solar system. Such a requirement can be relaxed on galactic scales and above, where data still gives the freedom to find possible signatures of their presence.

To make predictions to compare with observations coming from galactic and clusters scales (i.e. in the non-linear regime of cosmological evolution), cosmological N-body simulations are needed, for which codes must be developed that can solve for the scalar field. In this work we presented a new implementation in an N-body code of scalar-tensor theories of gravity that includes screening mechanisms. The code is based on the existing code \tt{RAMSES}, to which we have added a non-linear multigrid solver that can treat a large class of scalar tensor theories of modified gravity. We presented details of the implementation and the tests that we made to it.  
 
 As application of the new code, we studied the influence of two particular modified gravity theories, the symmetron and a chameleon-$f(R)$ gravity, on the shape of group and cluster-sized dark matter haloes. Using static calculations,  \citet[][]{2013PhRvL.110o1104L} show that the shape of the scalar fields follows the density distribution more closely than the gravitational potential, which should increase the ellipticity of the clusters, hopefully beyond observational limits.  If this is the case, the ellipticity of clusters provides a way to constraint the parameter space of scalar-tensor theories.  The time evolution was a missing ingredient in the \citet[][]{2013PhRvL.110o1104L} calculations.  Here we extended this work, including the time evolution in the dark matter component, and found that indeed the ellipticity of the clusters in the scalar-tensor simulations is higher, with exception of the most extreme $f(R)$ model studied. The predictions that we obtained for the $f(R)$ model were obtained in a different region of the parameter space studied in \citet[][]{2013PhRvL.110o1104L}, so it is not possible to make a direct comparison.  The main difference that we found with respect to the \citet[][]{2013PhRvL.110o1104L} expectations is that the $f(R)$ haloes can actually be slightly more spherical than $\Lambda$CDM ones when the fifth force is very long ranged. Further study of the formation history of the haloes is needed to fully understand this effect.

In the symmetron case we found results that are consistent with previous analytic estimations that exist in the literature.  Furthermore, we studied the mass dependence of this effects and found that different regions of the parameter space of this model give different dependencies.  This could help to distinguish between models once the influence of gas dynamics has been understood (not only in this models, but also within standard gravity), and accurate predictions can be made.
 
It is important to note that our results are based on a different quantity than studied in \citet[][]{2013PhRvL.110o1104L}.  There, the shape of the X-ray component was studied by assuming static dark matter haloes and hydro-static equilibrium, while here we studied the influence of the cosmological evolution in the shape of the underlying dark matter haloes.  The extension of these predictions to the shape of the gas component is beyond the scope of this work but will be performed elsewhere.  The impact that these results have on strong lensing statistics is left for future work.  What we can say at present is that the increase in the ellipticity will increase the probability of finding strong lenses, and thus, act in the right direction for solving the problem with the lensing statistics that the $\Lambda$CDM paradigm has \citep[see ][for a review of this topic]{2013SSRv..tmp...54M}.

\begin{acknowledgements}
C.L., D.F.M., and H.A.W thank the Research Council of Norway FRINAT grant 197251/V30. D.F.M. is also partially supported by projects CERN/FP/123618/2011 and CERN/FP/123615/2011.  The simulations were performed on the NOTUR Clusters \texttt{HEXAGON} and \texttt{STALLO}, the computing facilities at the Universities of Bergen and Troms{\o}, Norway.  We thanks Mikjel Thorsrud and Iain Brown for useful discussions.
\end{acknowledgements}

\bibliography{references}

\begin{thebibliography}{48}
\expandafter\ifx\csname natexlab\endcsname\relax\def\natexlab#1{#1}\fi

\bibitem[{{Amendola} {et~al.}(2012){Amendola}, {Appleby}, {Bacon}, {Baker},
  {Baldi}, {Bartolo}, {Blanchard}, {Bonvin}, {Borgani}, \&
  et~al.}]{2012arXiv1206.1225A}
{Amendola}, L., {Appleby}, S., {Bacon}, D., {et~al.} 2012, ArXiv:1206.1225
  [astro-ph.CO]

\bibitem[{{Baldi}(2012{\natexlab{a}})}]{baldi_de_sim}
{Baldi}, M. 2012{\natexlab{a}}, Physics of the Dark Universe, 1, 162

\bibitem[{{Baldi}(2012{\natexlab{b}})}]{2012MNRAS.422.1028B}
{Baldi}, M. 2012{\natexlab{b}}, \mnras, 422, 1028

\bibitem[{{Baldi} {et~al.}(2010){Baldi}, {Pettorino}, {Robbers}, \&
  {Springel}}]{baldi_coupled_quintessence_code}
{Baldi}, M., {Pettorino}, V., {Robbers}, G., \& {Springel}, V. 2010, \mnras,
  403, 1684

\bibitem[{{Behroozi} {et~al.}(2013){Behroozi}, {Wechsler}, \&
  {Wu}}]{2011arXiv1110.4372B}
{Behroozi}, P.~S., {Wechsler}, R.~H., \& {Wu}, H.-Y. 2013, \apj, 762, 109

\bibitem[{{Bertschinger}(1995)}]{1995astro.ph..6070B}
{Bertschinger}, E. 1995, arXiv:9506070 [astro-ph]

\bibitem[{{Brandt}(1977)}]{Brandt77}
{Brandt}, A. 1977, Math. of Comp., 31, 333

\bibitem[{{Brax} {et~al.}(2012){Brax}, {Davis}, {Li}, {Winther}, \&
  {Zhao}}]{2012JCAP...10..002B}
{Brax}, P., {Davis}, A.-C., {Li}, B., {Winther}, H.~A., \& {Zhao}, G.-B. 2012,
  \jcap, 10, 2

\bibitem[{{Brax} {et~al.}(2013){Brax}, {Davis}, {Li}, {Winther}, \&
  {Zhao}}]{nbody_chameleon}
{Brax}, P., {Davis}, A.-C., {Li}, B., {Winther}, H.~A., \& {Zhao}, G.-B. 2013,
  \jcap, 4, 29

\bibitem[{Brax {et~al.}(2008)Brax, van~de Bruck, Davis, \& Shaw}]{brax}
Brax, P., van~de Bruck, C., Davis, A.-C., \& Shaw, D.~J. 2008, Phys.Rev., D78,
  104021

\bibitem[{{Clifton} {et~al.}(2012){Clifton}, {Ferreira}, {Padilla}, \&
  {Skordis}}]{2012PhR...513....1C}
{Clifton}, T., {Ferreira}, P.~G., {Padilla}, A., \& {Skordis}, C. 2012,
  \physrep, 513, 1

\bibitem[{{Davis} {et~al.}(2012){Davis}, {Li}, {Mota}, \&
  {Winther}}]{2012ApJ...748...61D}
{Davis}, A.-C., {Li}, B., {Mota}, D.~F., \& {Winther}, H.~A. 2012, \apj, 748,
  61

\bibitem[{{Dubinski} \& {Carlberg}(1991)}]{1991ApJ...378..496D}
{Dubinski}, J. \& {Carlberg}, R.~G. 1991, \apj, 378, 496

\bibitem[{{Eriksen} {et~al.}(2004){Eriksen}, {Hansen}, {Banday}, {G{\'o}rski},
  \& {Lilje}}]{2004ApJ...605...14E}
{Eriksen}, H.~K., {Hansen}, F.~K., {Banday}, A.~J., {G{\'o}rski}, K.~M., \&
  {Lilje}, P.~B. 2004, \apj, 605, 14

\bibitem[{Gibou {et~al.}(2001)Gibou, Fedkiw, tien Cheng, \&
  Kang}]{Gibou_asecond-order-accurate}
Gibou, F., Fedkiw, R., tien Cheng, L., \& Kang, M. 2001, J. Comput. Phys, 205

\bibitem[{{Gr{\"o}nke} {et~al.}(2013){Gr{\"o}nke}, {Llinares}, \&
  {Mota}}]{2013arXiv1307.6994G}
{Gr{\"o}nke}, M.~B., {Llinares}, C., \& {Mota}, D.~F. 2013, arXiv:1307.6994
  [astro-ph.CO]

\bibitem[{{Guillet} \& {Teyssier}(2011)}]{2011JCoPh.230.4756G}
{Guillet}, T. \& {Teyssier}, R. 2011, Journal of Computational Physics, 230,
  4756

\bibitem[{{Hansen} {et~al.}(2009){Hansen}, {Banday}, {G{\'o}rski}, {Eriksen},
  \& {Lilje}}]{2009ApJ...704.1448H}
{Hansen}, F.~K., {Banday}, A.~J., {G{\'o}rski}, K.~M., {Eriksen}, H.~K., \&
  {Lilje}, P.~B. 2009, \apj, 704, 1448

\bibitem[{{Hinterbichler} \& {Khoury}(2010)}]{2010PhRvL.104w1301H}
{Hinterbichler}, K. \& {Khoury}, J. 2010, Physical Review Letters, 104, 231301

\bibitem[{{Hu} \& {Sawicki}(2007)}]{2007PhRvD..76f4004H}
{Hu}, W. \& {Sawicki}, I. 2007, \prd, 76, 064004

\bibitem[{Khoury \& Weltman(2004)}]{cham}
Khoury, J. \& Weltman, A. 2004, Phys.Rev.Lett., 93, 171104

\bibitem[{{Knebe} {et~al.}(2001){Knebe}, {Green}, \& {Binney}}]{Knebe01}
{Knebe}, A., {Green}, A., \& {Binney}, J. 2001, \mnras, 325, 845

\bibitem[{{Lau} {et~al.}(2011){Lau}, {Nagai}, {Kravtsov}, \&
  {Zentner}}]{2011ApJ...734...93L}
{Lau}, E.~T., {Nagai}, D., {Kravtsov}, A.~V., \& {Zentner}, A.~R. 2011, \apj,
  734, 93

\bibitem[{{Laureijs} {et~al.}(2011){Laureijs}, {Amiaux}, {Arduini},
  {Augu{\`e}res}, {Brinchmann}, {Cole}, {Cropper}, {Dabin}, {Duvet}, {Ealet},
  \& et~al.}]{2011arXiv1110.3193L}
{Laureijs}, R., {Amiaux}, J., {Arduini}, S., {et~al.} 2011, arXiv:1110.3193
  [astro-ph.CO]

\bibitem[{{Lee} \& {Baldi}(2012)}]{2012ApJ...747...45L}
{Lee}, J. \& {Baldi}, M. 2012, \apj, 747, 45

\bibitem[{{Li} \& {Barrow}(2011)}]{mlapmcode}
{Li}, B. \& {Barrow}, J.~D. 2011, \prd, 83, 024007

\bibitem[{Li {et~al.}(2011{\natexlab{a}})Li, Mota, \& Barrow}]{li2}
Li, B., Mota, D.~F., \& Barrow, J.~D. 2011{\natexlab{a}}, Astrophys.J., 728,
  109

\bibitem[{Li {et~al.}(2011{\natexlab{b}})Li, Mota, \& Barrow}]{li1}
Li, B., Mota, D.~F., \& Barrow, J.~D. 2011{\natexlab{b}}, Astrophys.J., 728,
  108

\bibitem[{{Li} {et~al.}(2013){Li}, {Zhao}, \& {Koyama}}]{dgp_code_durham}
{Li}, B., {Zhao}, G.-B., \& {Koyama}, K. 2013, \jcap, 5, 23

\bibitem[{{Li} {et~al.}(2012){Li}, {Zhao}, {Teyssier}, \&
  {Koyama}}]{2012JCAP...01..051L}
{Li}, B., {Zhao}, G.-B., {Teyssier}, R., \& {Koyama}, K. 2012, \jcap, 1, 51

\bibitem[{{Llinares}(2011)}]{llinares_thesis}
{Llinares}, C. 2011, PhD thesis, Univ. Groningen, ISBN: 978-90-367-4760-8
  http://dissertations.ub.rug.nl/faculties/science/2011/c.llinares

\bibitem[{{Llinares} {et~al.}(2008){Llinares}, {Knebe}, \&
  {Zhao}}]{2008arXiv0809.2899L}
{Llinares}, C., {Knebe}, A., \& {Zhao}, H. 2008, \mnras, 391, 1778

\bibitem[{{Llinares} \& {Mota}(2013{\natexlab{a}})}]{2013PhRvL.110p1101L}
{Llinares}, C. \& {Mota}, D.~F. 2013{\natexlab{a}}, Physical Review Letters,
  110, 161101

\bibitem[{{Llinares} \& {Mota}(2013{\natexlab{b}})}]{2013PhRvL.110o1104L}
{Llinares}, C. \& {Mota}, D.~F. 2013{\natexlab{b}}, Physical Review Letters,
  110, 151104

\bibitem[{{Llinares} {et~al.}(2009){Llinares}, {Zhao}, \&
  {Knebe}}]{2009ApJ...695L.145L}
{Llinares}, C., {Zhao}, H.~S., \& {Knebe}, A. 2009, \apjl, 695, L145

\bibitem[{{LSST Science Collaboration} {et~al.}(2009){LSST Science
  Collaboration}, {Abell}, {Allison}, {Anderson}, {Andrew}, {Angel}, {Armus},
  {Arnett}, {Asztalos}, {Axelrod}, \& et~al.}]{2009arXiv0912.0201L}
{LSST Science Collaboration}, {Abell}, P.~A., {Allison}, J., {et~al.} 2009,
  arXiv:0912.0201 [astro-ph.IM]

\bibitem[{{Martel} \& {Shapiro}(1998)}]{1998MNRAS.297..467M}
{Martel}, H. \& {Shapiro}, P.~R. 1998, \mnras, 297, 467

\bibitem[{{Meneghetti} {et~al.}(2013){Meneghetti}, {Bartelmann}, {Dahle}, \&
  {Limousin}}]{2013SSRv..tmp...54M}
{Meneghetti}, M., {Bartelmann}, M., {Dahle}, H., \& {Limousin}, M. 2013, \ssr

\bibitem[{{Oyaizu}(2008)}]{2008PhRvD..78l3523O}
{Oyaizu}, H. 2008, \prd, 78, 123523

\bibitem[{{Planck Collaboration} {et~al.}(2013{\natexlab{a}}){Planck
  Collaboration}, {Ade}, {Aghanim}, {Armitage-Caplan}, {Arnaud}, {Ashdown},
  {Atrio-Barandela}, {Aumont}, {Baccigalupi}, {Banday}, \&
  et~al.}]{planck_param}
{Planck Collaboration}, {Ade}, P.~A.~R., {Aghanim}, N., {et~al.}
  2013{\natexlab{a}}, arXiv:1303.5076 [astro-ph.CO]

\bibitem[{{Planck Collaboration} {et~al.}(2013{\natexlab{b}}){Planck
  Collaboration}, {Ade}, {Aghanim}, {Armitage-Caplan}, {Arnaud}, {Ashdown},
  {Atrio-Barandela}, {Aumont}, {Baccigalupi}, {Banday}, \&
  et~al.}]{sigma8_planck}
{Planck Collaboration}, {Ade}, P.~A.~R., {Aghanim}, N., {et~al.}
  2013{\natexlab{b}}, arXiv:1303.5080 [astro-ph.CO]

\bibitem[{{Puchwein} {et~al.}(2013){Puchwein}, {Baldi}, \&
  {Springel}}]{2013MNRAS.436..348P}
{Puchwein}, E., {Baldi}, M., \& {Springel}, V. 2013, \mnras, 436, 348

\bibitem[{{Schmidt}(2009)}]{schmidt_dgp_code}
{Schmidt}, F. 2009, \prd, 80, 043001

\bibitem[{{Shaw} {et~al.}(2006){Shaw}, {Weller}, {Ostriker}, \&
  {Bode}}]{2006ApJ...646..815S}
{Shaw}, L.~D., {Weller}, J., {Ostriker}, J.~P., \& {Bode}, P. 2006, \apj, 646,
  815

\bibitem[{{Teyssier}(2002)}]{2002A&A...385..337T}
{Teyssier}, R. 2002, \aap, 385, 337

\bibitem[{{Trottenberg} {et~al.}(2000){Trottenberg}, {Oosterlee}, \&
  {Scholler}}]{Trottenberg}
{Trottenberg}, U., {Oosterlee}, C., \& {Scholler}, A. 2000, {Multigrid}
  (Academic Press)

\bibitem[{{Wesseling}(1992)}]{Wesseling92}
{Wesseling}, P. 1992, {An Introduction to Multigrid Methods} (John Wiley and
  Sons Inc, |c1992, 2nd ed.)

\bibitem[{{Zhao} {et~al.}(2011){Zhao}, {Li}, \&
  {Koyama}}]{zhao_baojiu_forf_code}
{Zhao}, G.-B., {Li}, B., \& {Koyama}, K. 2011, \prd, 83, 044007

\end{thebibliography}

\end{document}